\input{aipcheck}

\documentclass[
    ,final            % use final for the camera ready runs
  ]
  {aipproc}

\layoutstyle{6x9}

\newcommand\nxt[1]  {\\\fnxt#1}

\def\calf         {{\cal F}}

\def\cali         {{\cal I}}

\def\caln         {{\cal N}}
\def\calo         {{\cal O}}

\def\calr         {{\cal R}}

\def\tr           {\mathop{\rm Tr}}

\def\a{\alpha}
\def\b{\beta}

\def\r{\rho}

\def\e{\epsilon}

\def\aa1{\phi}
\def\cc1{\psi}

\def\t{\tau}

\def\l{\lambda}
\def\ra{\Rightarrow}
\newcommand{\eq}{\begin{equation}}
\newcommand{\eqx}{\end{equation}}
\newcommand{\eqn}{\begin{eqnarray}}
\newcommand{\eqnx}{\end{eqnarray}}

\begin{document}

\title{On SUGRA description of boost-invariant conformal plasma at strong coupling
\footnote{Based on \cite{bbhj}.}}


\author{Alex Buchel}{
  address={Department of Applied Mathematics\\ University of Western Ontario\\
 London, Ontario N6A 5B7, Canada\\
Perimeter Institute for Theoretical Physics\\
Waterloo, Ontario N2J 2W9, Canada}
}

\begin{abstract}
We study string theory duals of the expanding boost invariant
conformal gauge theory plasmas at strong coupling. The dual
supergravity background is constructed as an asymptotic late-time
expansion, corresponding to equilibration of the gauge theory
plasma. The absence of curvature singularities in the first few orders
of the late-time expansion of the dual gravitational background
unambiguously determines the equilibrium equation of the state, and
the shear viscosity of the gauge theory plasma. While the absence of
the leading pole singularities in the gravitational curvature
invariants at the third order in late-time expansion determines the
relaxation time of the plasma, the subleading logarithmic singularity
can not be canceled within a supergravity approximation.  We comment
on the possible interpretations of this singularity.
\end{abstract}

\maketitle

The best understood example of gauge theory/string theory correspondence of Maldacena \cite{m1} is 
an equivalence between $\caln=4$ $SU(N)$ supersymmetric Yang-Mills (SYM) theory and type IIB string theory on 
$AdS_5\times S^5$. In the planar ('t Hooft) limit and at weak effective coupling $\lambda\equiv g_{YM}^2 N\ll 1$,
perturbative gauge theory description is valid. When the  't Hooft coupling  is large, $\l\gg 1$, the gauge theory 
is best described by type IIB supergravity on   $AdS_5\times S^5$.

The main motivation of this work is to use string theory (in a context of Maldacena correspondence) as a guiding principle 
in constructing {\it Non-equilibrium Quantum Field Theory}. As a first step in this direction we would like to use string 
theory to formulate dissipative relativistic theory of conformal fluids.

In the next section we describe boost-invariant expansion of  conformal fluids from the phenomenological 
perspective. We start with the ideal CFT fluid dynamics, then discuss the first-order dissipative CFT fluid dynamics.
We explain the reason of going beyond the first-order hydrodynamics and outline the M{\"u}ller-Israel-Stewart 
transient theory \cite{m,is}.  In section II we review aspects of gauge theory/string theory correspondence 
relevant for the discussion of a CFT plasma expansion. Specifically, we emphasize nonsingularity of the background 
geometry as a guiding principle to determine correct physics. In section III we review Janik-Peschanski proposal\footnote{See also 
\cite{j12,j2,j3}.} \cite{j1}
for string theory dual to boost-invariant plasma expansion. We discuss successes of the proposal 
(correct equation of state, shear viscosity, relaxation time), and  explain singularities 
in the supergravity approximation. We conclude in section IV with the possible interpretation of the singularities 
and outline future directions.

\section{I:\ Boost-invariant expansion of the conformal fluids}
Consider expansion of a CFT fluid (a gauge theory plasma) in the boost-invariant frame\footnote{This is expected to be 
a correct description of a central region of QGP produced in ultra-relativistic collisions of heavy nuclei \cite{bj}.}. 
We convert Minkowski frame 
\[
ds_4^2=-dx_0^2+dx_\perp^2+dx_3^2
\]
into a frame with the boost-invariance along $x_3$ direction
\[
x_0=\t\ \cosh y\,,\qquad x_3=\t\ \sinh y 
\]
\[
ds_4^2=-d\t^2+\t^2\ dy^2+dx_\perp^2\,.
\]
We further assume 
\[
\e=\e(\t)\,,\qquad P=P(\t)\,,
\]
for a local energy density $\e$ and a pressure $P$ in a fluid.

The stress-energy tensor of an ideal fluid takes form 
\[
T_{\mu\nu}\equiv T_{\mu\nu}^{equilibrium}=(\e+P)\ u_\mu u_\nu+p\ \eta_{\mu\nu}\,,
\]
where $u^{\mu}$ is a local 4-velocity of the fluid, $u^2=-1$. Scale invariance of a CFT fluid implies 
\[
T_{\mu}^\mu=0\qquad \Rightarrow\qquad \e=3 p\,,
\]
so that the stress-energy conservation  in the boost-invariant frame takes a simple form
\[
\nabla_\mu T^{\mu\nu}=0\qquad \Rightarrow\qquad \frac{d\e}{d\t}=-\frac 43\frac {\e}{\t}\,.
\]
Above equation describes evolution of the energy density of the expanding fluid at equilibrium.
We find 
\[
\e\propto \t^{-4/3}\,.
\]
Since energy density is the only scale in a conformal fluid, dimensional arguments readily establish 
proper time scaling for the fluid temperature $T$,  entropy density $s$,  shear viscosity $\eta$,
and  relaxation time $\t_\pi$:  
\[
T\propto \e^{1/4}\propto \t^{-1/3}\,,\qquad \eta\propto s\propto T^3\propto \t^{-1}\,,\qquad \t_\pi\propto T^{-1}\propto \t^{1/3}\,.
\]

In the first-order dissipative CFT fluid dynamics the stress-energy tensor is phenomenologically modified from its equilibrium value
by a symmetric, transverse and traceless tensor $\t_{\mu\nu}$ constructed from the first derivatives of the 
4-velocity $u^\mu$:   
\[
T_{\mu\nu}=T_{\mu\nu}^{equilibrium}+\t_{\mu\nu}\,,\ \ \t^{\mu\nu}=-\eta\left(
\Delta^{\mu\a}\Delta^{\nu\b}(\nabla^\mu u^\nu+\nabla^\nu u^\mu)-\frac 23 \Delta^{\mu\nu}\Delta^{\a\b} \nabla_\a u_\b
\right)\,,
\]
where $\eta$ is the shear viscosity of the fluid\footnote{Scale invariance  forbids nonzero 
bulk viscosity for a conformal fluid.}, and 
\[
\Delta^{\mu\nu}\equiv \eta^{\mu\nu}+u^\mu u^\nu\,.
\]
In this case, the stress-energy conservation in the boost-invariant frame leads to 
\[
\frac{d\e}{d\t}=-\frac 43\frac {\e}{\t}+\frac{4\eta}{3\t^2}\,.
\]
From the equilibrium scaling for the boost-invariant expansion we see that  the viscous correction becomes subdominant 
as $\t\to \infty$. Indeed, 
\[
\frac \e\t\bigg|_{equilibrium}\propto \frac{\t^{-4/3}}{\t}\propto \t^{-7/3}\,,\qquad \frac {\eta}{\t^2}
\bigg|_{equilibrium}\propto \frac{\t^{-1}}{\t^2}
\propto \t^{-9/3}\,.
\]
Thus we expect approach to equilibrium in the boost-invariant frame to correspond to the late-time dynamics.

Why do we need to consider  dissipative hydrodynamics beyond the first-order? 
It turns out that the first-order hydrodynamics 
allows for an acausal signal propagation. 
Though we do not cover causality violation in the first-order hydrodynamics here\footnote{See \cite{mur} for a detailed discussion of this.},
such a violation would be clear once we discuss the second-order (causal) dissipative CFT fluid dynamics of M{\"u}ller-Israel-Stewart
(MIS)
\cite{m,is}. Basic equations of MIS theory in the boost-invariant frame for a CFT fluid 
take form 
\[
0=\frac{d\e}{d\t}+\frac{\e+P}{\t}-\frac{1}{\t}\Phi\,,
\]
\[
0=\frac{d\Phi}{d\t}+\frac{\Phi}{\t_\pi}+\frac 12 \Phi\left(\frac 1\t+ \frac{1}{\b_2}T\frac{d}{d\t}\left(\frac{\b_2}{T}\right)\right)
-\frac 23 \frac 1\b_2\frac 1\t\,,
\]  
where $\t_\pi$ is the relaxation time, $\Phi$ is related to the dissipative part of the stress-energy, and 
\[
\b_2=\frac{\t_\pi}{2\eta}\,.
\]
From the equilibrium scaling, $\t\to \infty$ limit corresponds effectively to $\t_\pi\to 0$ and the second-order hydrodynamics of MIS 
reduces to the first-order hydrodynamics. Since in this limit the relaxation is effectively instantaneous, it is not surprising
that causality in the first-order hydrodynamics is violated.

We would like to apply the second-order hydrodynamics to $\caln=4$ $SU(N)$ SYM plasma in the boost-invariant frame.
Again, since local temperature is the only scale in the problem, we expect
\[
\e(\t)=\frac 38\pi^2 N^2 T(\t)^4\,,\qquad P(\t)=\frac 13 \e(\t)\,,\qquad \eta(\t)=A\ s(\t)=A\ \frac 12 \pi^2 N^2 T(\t)^3
\]  
\[
\t_\pi(\t)=r\ \t_{\pi}^{Boltzmann}(\t)=r\ \frac{3\eta(\t)}{2P(\t)}
\]
where $A$ is the ratio of shear viscosity to entropy density and $r$ is the relaxation time in units of Boltzmann relaxation time.
From MIS equations as $\t\to \infty$:
\[
T(\t)=\frac{\Lambda}{\t^{1/3}}\left(1+\sum_{k=1}^\infty \frac{t_k}{(\Lambda\t^{2/3})^k}\right)\,,\qquad
\Phi(\t)=\frac 23\pi^2 N^2 A \frac{\Lambda^3}{\t^2}\left(1+\sum_{k=1}^\infty\frac{f_k}{(\Lambda\t^{2/3})^k}\right)\,,
\]
where $\Lambda$ is an arbitrary scale and 
\[
t_k=t_k(A,r)\,,\qquad f_k=f_k(A,r)\,.
\]
Notice that within the second-order MIS theory, $\caln=4$ hydrodynamics is uniquely specified by two dimensionless parameters: 
$\{A,r\}$. On the other hand, to completely specify a non-equilibrium state of a system 
one needs an {\it infinite} set of parameters. These additional 
parameters will show up as higher-order transport coefficients in constitutive relations between non-equilibrium part of the stress-energy 
tensor $\t^{\mu\nu}$ and higher-order gradients of the 4-velocity $u^\mu$. It is very difficult to specify such constitutive relations 
from the purely phenomenological approach, {\it i.e.}, along the lines of MIS theory. But it is precisely in this 
respect that the gauge/string theory correspondence of Maldacena can be useful\footnote{For recent developments in this direction
see \cite{r1,r2,r3,r4}.}.  

\section{II:\ Some aspects of gauge/theory string theory correspondence }

Maldacena correspondence is a duality between a gauge theory and a full string theory. However, the correspondence is 
useful when it is computationally tractable. Typically, this implies a truncation of the full string theory to 
its low-energy supergravity approximation. One has to be careful though: a particular supergravity truncation 
is not always consistent. Inconsistencies of the truncation are often reflected in singularities of the 
resulting supergravity backgrounds.  We can distinguish three main cases:
\nxt $(A)\Rightarrow$ in some cases singularities of the supergravity backgrounds are simply an indication that 
(further) Kaluza-Klein truncation of the supergravity is incorrect, and including a finite number of additional SUGRA modes 
(doing consistent KK truncation) one obtains a smooth geometry (for example: a black hole solution on the singular conifold 
with self-dual fluxes \cite{bh1,bh2});
\nxt $(B)\Rightarrow$ in some cases singularities of the supergravity backgrounds are expected to be resolved by including an 
infinite set of string theory $\a'$ corrections --- from the gauge theory perspective this would imply that an infinite set 
of gauge theory operators (of increasingly high dimension) would develop a vacuum expectation value at strong coupling;
\nxt $(C)\Rightarrow$ in some cases singularities of the supergravity backgrounds are not expected to 
be resolved within the full string theory, as this would falsify gauge/string correspondence --- string theory would predict a 
gauge theory phase which can not be realized physically (for example: a singularity of the Klebanov-Tseytlin geometry \cite{kt}
is not expected to be resolved in string theory preserving both the supersymmetry and the chiral symmetry). 

A singularity in a given supergravity truncation would imply correspondingly that: 
\nxt  $(A)\Rightarrow$ a consistency of supergravity would determine additional operators on the gauge theory side that would develop 
a VEV at strong coupling (for a finite temperature Klebanov-Tseytlin background, SUGRA is smooth once a $U(1)$ fiber inside 
$T^{1,1}$ is warped \cite{bh2,bh3} $\Leftrightarrow$ a dim-6 operators of the thermal gauge theory plasma develops a VEV \cite{bh4,bh5}); 
\nxt  $(B)\Rightarrow$ SUGRA truncation is not useful;
\nxt  $(C)\Rightarrow$ a phase of the gauge theory with prescribed symmetries simply does not exists 
(a singular Klebanov-Tseytlin solution is replaced with a smooth Klebanov-Strassler solution \cite{ks}
where the chiral symmetry is broken).

\section{III:\ Janik-Peschanski proposal for string theory dual to boost invariant expansion}

Given the symmetries of the expanding boost-invariant $\caln=4$ SYM plasma, the most general parity invariant truncation of the dual 
type IIB SUGRA  takes form 
\[
ds_{10}^2=e^{-2\a(\t,z)}\left\{\frac {1}{z^2}\left[-e^{2a(\t,z)}d\t^2+e^{2b(\t,z)}\t^2dy^2+e^{2c(\t,z)}dx_\perp^2\right]
+\frac{dz^2}{z^2}\right\}
\]
\[
+e^{6/5\a(\t,z)}\left(dS^5\right)^2
\]
for the Einstein frame metric;
\[
F_5=\calf_5+\star \calf_5\,,\qquad \calf_5=-4Q\ \omega_{S^5}\,,\qquad \phi=\phi(\t,z)\,,
\]
for the 5-form ($Q$ is a constant related to the rank of the gauge group) and the dilaton. 
We set 
\[
Q=1\qquad \Leftrightarrow\qquad R_{AdS_5}=1\,.
\]
Asymptotically as $z\to 0$
\[
\{a,b,c,\a,\phi\}\rightarrow 0\,,
\]
to insure that the boundary geometry is that of the boost-invariant frame; however, we assume that 
\[
a(\t,z)\sim \calo(z^4)\ne 0\,,
\]
so that the local energy density in the expanding plasma is nonzero.

We proceed then as follows:

\noindent $\Rightarrow$ we try to construct a nonsingular geometry everywhere in the bulk, subject to the above boundary conditions;

\noindent $\ra$ we evaluate the stress-energy tensor one-point correlation function 
\[
\left\langle T_{\mu\nu}(\t) \right\rangle =\frac{N_c^2}{2\pi} \lim_{z\to 0}
\frac{g_{\mu\nu}^{(5)}(\t)-\eta_{\mu\nu}}{z^4}\,;
\]

\noindent $\ra$ we extract from $\left\langle T_{\mu\nu}(\t) \right\rangle$
\[
\e(\t)\,,\qquad p(\t)\,,
\]
and interpret results in the framework of the dissipative relativistic fluid dynamics.

We saw before that near-equilibrium hydrodynamics corresponds to the late-time asymptotic expansion of the 
boost-invariant conformal plasma. Thus, it appears natural to assume \cite{j1,j2,j3}
\[
a(\t,z)=a\left(\t,v\equiv \frac{z}{\t^s}\right)\,,
\]
as well as for the remaining SUGRA modes; then, study background geometry as asymptotic expansion in $\t$ 
while keeping the scaling variable $v$ finite.

To leading order as $\t\to \infty$, the absence of singularities in 
\[
\cali^{[2]}\equiv \calr_{\mu\nu\r\l}\calr^{\mu\nu\r\l}\,,\qquad v^4\to 3_- \,,
\]
requires 
\[
s=\frac 13\,.
\]

Given the value of $s$, the asymptotic expansion for the warp factors of the 5-dim geometry takes form
\[
a(\tau,v) = a_{0}(v) + \frac{1}{\tau^{2/3}} a_{1}(v) + \frac{1}{\tau^{4/3}} 
a_{2}(v) + \frac{1}{\tau^{2}} a_{3}(v)+\calo(\t^{-8/3})\,,
\]
\[
b(\tau,v) = b_{0}(v) + \frac{1}{\tau^{2/3}} b_{1}(v) 
+ \frac{1}{\tau^{4/3}} b_{2}(v) + \frac{1}{\tau^{2}} b_{3}(v)+\calo(\t^{-8/3})\,,
\]
\[
c(\tau,v) = c_{0}(v) + \frac{1}{\tau^{2/3}} c_{1}(v) 
+ \frac{1}{\tau^{4/3}} c_{2}(v) + \frac{1}{\tau^{2}} c_{3}(v)+\calo(\t^{-8/3})\,,
\]
which leads to the late-time expansion of the Riemann tensor squared
\[
\cali^{[2]}
=\cali^{[2]}_0(v)+\frac{1}{\t^{2/3}} \cali^{[2]}_1(v)+\frac{1}{\t^{4/3}} \cali^{[2]}_2(v)+\frac{1}{\t^{2}} \cali^{[2]}_3(v)+\calo(\t^{-8/3})\,.
\]

Let's assume first 
\[
\a(\t,v)\equiv 0\,,\qquad \phi(\t,v)\equiv 0 \,,
\]
which on the gauge theory side implies that neither $\left\langle \tr F^2(\t) \right\rangle$ (dual to a 
dilaton) nor the dim-8 operator (dual to a SUGRA scalar $\a$) develop a VEV.
We emphasize that this is an assumption which might or might not  be correct --- we use nonsingularity condition 
of the dual string (supergravity) description  to test it.

We find (up to the second subleading order)
\[
\e(\t)=\left(\frac{N^2}{2\pi^2}\right)\ \frac{1}{\t^{4/3}}\biggl\{1-\frac{2\eta_0}{\t^{2/3}}+\left(\frac{10}{3}\eta_0^2+\frac{C}{36}
\right)\frac{1}{\t^{4/3}}+\cdots\biggr\}\,.
\]
Matching the gauge theory expansion for the energy density  with that of the dual gravitational description
we find
\[
\Lambda=\frac{\sqrt{2}}{3^{1/4}\pi }\,,\qquad A=\frac{3^{3/4}}{2^{3/2}\pi}\ \eta_0\,,
\qquad r=-\frac{11}{18}-\frac{1}{108}\ \frac{C}{\eta_0^2}\,.
\]
As we alluded to before, further expansion on the SUGRA side will {\it define} higher order dissipative relativistic dynamics. 

For generic values of $\{\eta_0,C\}$: 
\[
\biggl\{\cali^{[2]}_2(v)\,,\cali^{[2]}_3(v)\biggr\}=\calo\left(\frac{1}{(3-v^4)^{4}}\right)\,,\qquad v^4\to 3_-\,.
\]
Tuning 
\[
\eta_0=\frac{1}{2^{1/2}3^{3/4}}\,,\qquad C=2\sqrt{3}\ \ln2 -\frac{17}{\sqrt{3}}\,,
\]
{\it all} pole singularities in $\{\cali^{[2]}_2(v)\,,\cali^{[2]}_3(v)\}$ are removed.
Mapping above values on the gauge theory side we find the ratio of the shear viscosity to the entropy density 
and the plasma relaxation time (within MIS phenomenological theory)
\[
A=\frac{1}{4\pi}\,,\qquad r=\frac 13(1-\ln 2)\,.
\]
Notice that the shear viscosity agrees\footnote{Such an agreement is lost once 
finite 't Hooft coupling corrections are taken into account \cite{fc1,fc2,fc3}. 
The latter suggests that the commonly used $\calo(\a'^3)$ string theory effective action is incomplete (see also \cite{s}).} 
with earlier computations using equilibrium correlation 
functions \cite{pss}.

However:
\[
\cali^{[2]}_3={\rm finite}\ + \left(8\ 2^{1/2}\ 3^{3/4}\right)\ \ln(3-v^4)\,,\qquad v\to 3^{1/4}_-\,.
\]
Thus it appears inconsistent to set $\a$ and/or the dilaton to zero; in fact weak coupling analysis suggests that 
there are instabilities in expanding plasma generating VEVs of various operators \cite{ins1}, in particular 
$\left\langle \tr F^2(\t) \right\rangle$.
A careful analysis shows that without introducing additional pole curvature singularities one can turn on 
only the $\a$ mode to relevant order
\[
\a(\t,v)=\frac{1}{\t^2}\ \a_3(v)+\calo\left(\t^{-8/3}\right)\,,
\]
\[
\a_3=\a_{3,0}\left( \left(\frac{1}{96v^4}+\frac{v^4}{864}\right)\ \ln\frac{3+v^4}{3-v^4}-\frac{1}{144}\right)\,,
\]
where $\a_{3,0}$ is a normalizable mode related to the VEV of the dim-8 operator in $\caln=4$ SYM plasma.

We find:
\[
\cali^{[2]}_3={\rm finite}\ + \left(8\ 2^{1/2}\ 3^{3/4}+\frac{14}{3}\ \a_{3,0}\right)\ \ln(3-v^4)\,,\qquad v\to 3^{1/4}_-\,,
\]
but
\[
\calr_{\mu\nu}\calr^{\mu\nu}={\rm finite}+\frac{1}{\t^2}\ \frac{40}{3}\ \a_{3,0}\ \ln(3-v^4)\,,\qquad v\to 3^{1/4}_-\,.
\]
So, logarithmic singularity can not be canceled within the SUGRA approximation.

We considered another model of a CFT plasma (specifically, Klebanov-Witten plasma \cite{kw}) which has an additional SUGRA mode 
and showed that logarithmic singularities both in 
\[
\calr_{\mu\nu}\calr^{\mu\nu}\qquad {\rm and}\qquad \calr_{\mu\nu\r\l}\calr^{\mu\nu\r\l}
\]
at the third subleading order can be canceled (Ricci scalar is  nonsingular).
However, new logarithmic singularities appear at the third order in higher curvature
invariants such as
\[
\calr_{\mu_1\nu_1\l_1\r_1} \calr^{\mu_1\nu_1\l_2\r_2} \calr_{\mu_2\nu_2}\ ^{\l_1\r_1}\calr^{\mu_2\nu_2}\ _{\l_2\r_2}\,,
\] 
 as well as  logarithmic
singularities with  different coefficients in 
\[
(\calr_{\cdot\cdot\cdot\cdot})^8\,,\qquad  (\calr_{\cdot\cdot\cdot\cdot})^{16}\,,
\] 
and so on.
Thus, we conclude that one needs an infinite set of fields to cancel singularities in gravitational description, corresponding to 
an infinite set of gauge invariant operators developing a VEV during boost-invariant expansion.

\section{IV:\ Interpretation of singularities and future directions}

We attempted to construct a string theory  dual to strongly coupled conformal expanding
plasmas in the Bjorken regime. In order to have computational control we truncated the full string theory to supergravity
approximation, and focused on the well-established examples of the gauge/string dualities:
we considered $\caln=4$ SYM  and superconformal Klebanov-Witten  gauge theories. We used
nonsingularity of the dual gravitational backgrounds as a guiding principle to identify gauge theory
operators that would develop a vacuum expectation value during boost-invariant expansion of the plasma. 
Truncation to
a supergravity sector of the string theory (along with parity invariance in the Bjorken frame) severely restricts a
set of such operators. In the case of the $\caln=4$  SYM, there are only two such gauge invariant operators, while
for the Klebanov-Witten plasma one has an additional operator. We constructed supergravity dual as a late-time
asymptotic expansion and demonstrated that the gravitational boundary stress energy tensor expectation value has exactly the
same asymptotic late-time expansion as predicted by M\"uller-Israel-Stewart theory of transient relativistic kinetic  theory for the
boost-invariant expansion. As an impressive success of this approach, one recovers by requiring
nonsingularity of the background geometry at leading and the first three subleading orders\footnote{ At the third
subleading order, logarithmic singularities of the metric curvature invariants remain.}  the
equation of state for the plasma and its shear viscosity, in agreement with values extracted from the
equilibrium correlation functions. Unfortunately, we showed that logarithmic singularity in the
background geometry can not be canceled within the supergravity approximation. Moreover, given that the singularities
appear to persist in arbitrary high order metric curvature invariants, we suspect that relaxing the constraint of parity invariance in the
Bjorken regime would not help. Indeed, relaxing parity invariance would allow for only finite number of additional
(massive) supergravity modes, which, as an example of Klebanov-Witten plasma demonstrates, would only allow to cancel
logarithmic singularities in a finite number of additional metric curvature invariants.

Coming back to our classification of singularities in SUGRA truncations of the full string theory we see that for the 
string theory duals of the boost-invariant CFT plasmas expansion  
\nxt $(A)\ra$ is not realized;
\nxt $(B)\ra$ can be realized; in this case, though SUGRA truncation is inconsistent, maybe the requirement of the cancellation of the pole singularities 
at low orders is a correct prescription to extract the second order transport coefficients (which are of relevance to RHIC);
tantalizingly, we see hints of the universality of the relaxation time --- further study of  
non-conformal models is needed;
\nxt $(C)\ra$ can be realized; in this case a SUGRA singularity might be an indication of a genuine singularity in the full string theory description --- 
here, it is important to search for onset of instabilities/turbulence  in the expanding plasma.

\begin{theacknowledgments}

I would like to thank the Organizers of "Ten Years of AdS/CFT" conference for arranging an excellent meeting
in beautiful Buenos Aires. 
I would like to thank Paolo Benincasa, Michal Heller and Romuald Janik for collaboration on \cite{bbhj}.
My research at Perimeter Institute is supported in part by the Government
of Canada through NSERC and by the Province of Ontario through MRI.
I gratefully acknowledges further support by an NSERC Discovery
grant and support through the Early Researcher Award program by the
Province of Ontario.

\end{theacknowledgments}

\end{document}